\documentclass[twocolumn,showpacs,preprintnumbers,amsmath,amssymb,pra]{revtex4}

\usepackage{graphicx}
\usepackage{bm}
\usepackage{dcolumn}
\usepackage{array}

\newcommand{\triplet}{a \, ^3\Sigma_u^+}
\newcommand{\singlet}{X \, ^1\Sigma_g^+}
\newcommand{\tripletex}{1 \, ^3\Sigma_g^+}
\newcommand{\vib}{v}
\newcommand{\qme}{|\textrm{e}\rangle}
\newcommand{\qmf}{|\textrm{i}\rangle}
\newcommand{\qmv}{|v\rangle}
\newcommand{\clg}{1_{\textrm{g}}}
\newcommand{\cog}{0_{\textrm{g}}^-}
\newcommand{\wn}{$\mathrm{cm}^{-1}$}

\begin{document}

\title{Hyperfine, rotational, and vibrational structure
of the $\triplet$ state of $^{87}$Rb$_2$}

\author{C. Strauss $^{*1,2}$}
\author{T. Takekoshi $^{*1}$}
\author{F. Lang $^{1}$}
\author{K. Winkler $^{1}$}
\author{R. Grimm $^{1}$}
\author{J. Hecker Denschlag  $^{1,2}$}
\affiliation{$^{1}$Institut f\"ur Experimentalphysik und Zentrum
f\"ur Quantenphysik,\\ Universit\"at Innsbruck, 6020 Innsbruck, Austria\\
$^{2}$Institut f\"ur Quantenmaterie,\\ Universit\"at Ulm, 89069
Ulm, Germany }
\author{E. Tiemann $^{3}$}
\affiliation{$^{3}$Institute of Quantum Optics,\\  Leibniz
Universit\"at Hannover, 30167 Hannover, Germany}

\date{\today}

\begin{abstract}
We have performed high-resolution two-photon dark-state
spectroscopy of an ultracold gas of $^{87}$Rb$_2$ molecules in the
$\triplet$ state at a magnetic field of about 1000\,G. The
vibrational ladder as well as the hyperfine and low-lying
rotational structure is mapped out.  Energy shifts in the spectrum
are observed due to singlet-triplet mixing at binding energies as
deep as a few hundred GHz$\times h$. This information together
with data from other sources is used to optimize the potentials of
the $\triplet$ and $\singlet$ states in a coupled-channel
model. We find that the hyperfine structure depends weakly on the
vibrational level. This provides a possible explanation for
inaccuracies in recent Feshbach resonance calculations.
\end{abstract}

%     cooling mol.;   trap molec.;   atom in lattice      Quant opt.
% PACS, the Physics and Astronomy Classification Scheme.

\maketitle

\section{Introduction}

Recently, optical schemes have been developed to selectively
produce cold and dense samples of deeply bound molecules in well
defined quantum states \cite{Dan08, La08, Ni08} (see also
\cite{Vit08, Dei08}). This has opened up new possibilities for
cold collision experiments, ultracold chemistry \cite{Sta06,
Kre05, Kre08}, and for testing fundamental laws via precision
spectroscopy \cite{Ze08, DeM08, Ch09}. For such future experiments
it is mandatory that the location and properties of the available
molecular quantum states are well known and understood.

\begin{figure}[tpb]
\centering
\includegraphics{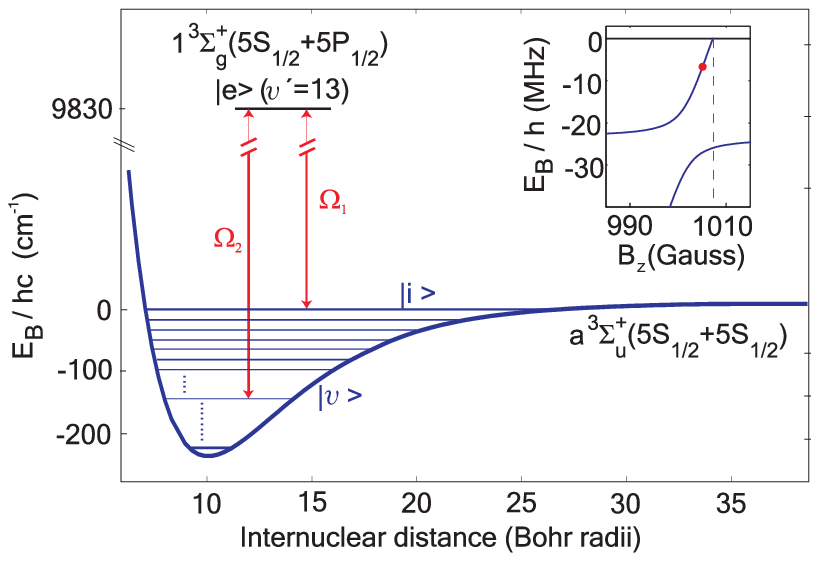}
\caption{Dark-state spectroscopy scheme for the $^{87}$Rb$_2$
$\triplet$ potential. Lasers 1 and 2 couple the molecular levels
$\qmf$ and $\qmv$ to the excited level $\qme$ with Rabi
frequencies $\Omega_{1,2}$, respectively. Laser 1 is kept on
resonance while laser 2 can be tuned to any level of the
$\triplet$ potential. Inset: Bound state level of Feshbach
molecules  as a function of magnetic field $B_z$. The dashed line
gives the position of the Feshbach resonance. The dot marks the
Feshbach molecule state used in the experiments.}
\label{fig:potential}
\end{figure}

Until recently, the $\triplet$ triplet potential of the Rb$_2$
molecule has remained largely experimentally unexplored. Natural
samples of Rb$_2$ are normally found in their $\singlet$ singlet
state
 from which the lowest triplet state is somewhat
difficult to reach in an optical Raman process due to a change in
u/g symmetry. Here we report precision spectroscopy of
$^{87}$Rb$_2$ molecules in this lowest triplet state, $\triplet$
(5$S_{1/2}$ + 5$S_{1/2})$, where we resolve vibrational,
rotational, hyperfine and Zeeman structure with an accuracy as
high as 30\,MHz. We use dark-state spectroscopy where a gas of
weakly bound Feshbach molecules is irradiated by two laser beams.
Molecular losses, induced by one of the two lasers, are suppressed
when the second laser is tuned into resonance with a bound state
(see Fig.\,\ref{fig:potential}). Fitting a coupled-channel model
to our experimental data, we have constructed an accurate
Born-Oppenheimer potential for $\triplet$. This enables us to
calculate the wave functions of  triplet bound states as well as
their binding energies to 60\,MHz$\times h$ accuracy (where $h$ is
Planck's constant) over the whole manifold of vibrational and low
rotational (N$<$5) levels. Our theory and data agree to the extent
that further refinement of the theoretical model requires a
reduction in the experimental uncertainty.

Molecular spectroscopy with cold atomic gases goes back to the
beginnings of laser cooling \cite{Wei99,Jon06,Koh06}. An
experiment closely related to ours is the one by Araujo {\it et
al.} \cite{deA03} where the Na$_2$ triplet ground state was
explored using a magneto optical trap combined with two-color
photoassociation spectroscopy. Our spectroscopy also makes use of
a two-photon transition, but starts from cold Feshbach molecules
rather than from free atoms. Recent investigations of the
$\triplet$ potential of Rb$_2$ include the work of \cite{L06},
\cite{Be09} and \cite{Mu10}. Using one-color photoassociation of
the $\triplet$ potential of laser cooled $^{85}$Rb, reference
\cite{L06} put tight constraints on the position of the repulsive
wall of the $\triplet$ potential but was not able to resolve the
vibrational structure. The two other groups determined several
ro-vibrational levels using fluorescence spectroscopy \cite{Be09}
or pump-probe photoionization spectroscopy of Rb$_2$ formed on
helium nanodroplets \cite{Mu10}. Our work goes well beyond these
measurements as we fully resolve hyperfine, rotational and Zeeman
structure for almost all vibrational states. Highly precise data
of the asymptotic behavior of the coupled $\triplet$- $\singlet$
system is contained in the large set of observed Feshbach
resonances \cite{marte:2002, roberts:2001, erhard:2003, papp:2006}
and in the two-photon photoassociation measurements of four weakly
bound levels at zero magnetic field by \cite{wynar}.

This article is organized as follows: Section \ref{exp} presents
the experimental setup and typical dark state spectroscopy scans.
Section \ref{quantum} discusses the relevant quantum numbers of
our studies and the assignment of the observed lines. Section
\ref{ccmodel} is a short summary of the coupled-channel model and
the optimization procedure of the Born-Oppenheimer potentials.
Section \ref{vibmode} discusses the progression of the
substructure of the vibrational manifolds. We conclude the paper
with a summary and an outlook towards further experiments in
section \ref{conclusion}.

\section{Experimental setup and dark state spectroscopy}
\label{exp}

The starting point for our experiments is a 50\,$\mu$m-size pure
ensemble of $3\times10^4$ weakly bound Feshbach molecules which
have been produced from an atomic Bose-Einstein condensate of
$^{87}$Rb by ramping over a Feshbach resonance at a magnetic field
of 1007.4\,G (1\,G = 10$^{-4}$\,T) \cite{Vol03}. They are trapped
in the lowest Bloch band of a cubic 3D optical lattice with no
more than a single molecule per lattice site~\cite{Tha06}. The
lattice depth for the Feshbach molecules is 60\,$E_r$, where
$E_r=\pi^2\hbar^2/2m a^2$ is the recoil energy, with $m$ the mass
of the molecules and $a = 415.22\,$nm the lattice period. Such
deep lattices prevent the molecules from colliding with each
other, which suppresses collisional decay. We observe lifetimes of
a few hundred ms. After producing the Feshbach molecules, the
magnetic field is set to 1005.8\,G where the spectroscopy is
carried out. At this magnetic field, the binding energy of the
Feshbach molecules is 4.4(3)\,MHz$\times h$ (Fig.\,
\ref{fig:potential} inset).

Our dark state spectroscopy works as follows: The Feshbach
molecules in state $\qmf$ are irradiated by simultaneous
rectangular pulses from lasers 1 and 2. The pulses typically last
10\,$\mu$s with Rabi frequencies $\Omega_{1}$ and $\Omega_{2}$,
respectively (Fig.\,\ref{fig:potential}). We keep laser 1 resonant
with the $\qmf-\qme$ transition and at a power $I_1$ of about
0.1\,mW ($\Omega_{1}= 2 \pi \times $ 0.3\,MHz) such that in the
absence of laser 2 about half of the molecules are lost by
spontaneous emission from $\qme$. Laser 2 with its power $I_2$ (up
to a few hundred mW) is scanned. As long as laser 2 does not hit
an $\qme-\qmv$ resonance, laser 1 will continue to induce losses.
However, when an $\qme-\qmv$ resonance occurs, the initial state
$\qmf$ is projected onto a dark state $|\Psi_{dark}\rangle =
(\Omega_{2}\qmf - \Omega_{1}|v\rangle)/ \sqrt{\Omega_{1}^2+
\Omega_{2}^2}$. Molecules in this dark state are shielded from
excitation to the short-lived level $\qme$ \cite{La09}. This leads
to a suppression of molecular losses. After the lasers are
switched off we measure the number of molecules via a reverse
magnetic field sweep through the Feshbach resonance, dissociating
the remaining molecules in $\qmf$ into atoms which are detected by
absorption imaging. These measurements are destructive, and for
each point in a scan, a fresh sample of Feshbach molecules has to
be prepared.

\begin{figure}[tbp]
\centering
\includegraphics{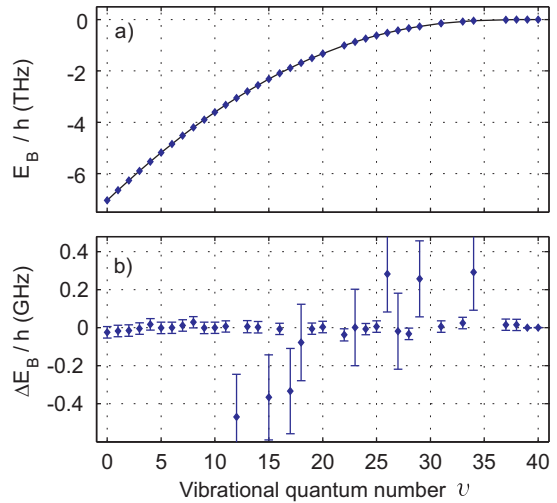}
\caption{(a) Binding energies $E_B(v)$ for the state $\triplet$,
where $v$ is the vibrational quantum number. The line is the
result of a coupled-channel model calculation after optimization
of the Born-Oppenheimer potential.  Five levels were not measured.
(b) Residues \emph{i.e.} difference between experimental data and
theory. Large error bars belong to early measurements without
simultaneous
wavemeter readings of both lasers.} %For details see text.
\label{fig:vibladder}
\end{figure}

The level $\qme$ has an excitation energy of about 295
\,THz$\times h$ with respect to $\qmf$ and a width $\Gamma$ =
$2\pi \times$ 8\,MHz. Laser 1, a grating-stabilized diode laser,
is Pound-Drever-Hall locked to a cavity which is in turn locked to
an atomic $^{87}$Rb-line. Laser 2, a Ti:Sapphire laser, is
free-running and typically drifts over a frequency range of a few
MHz within seconds. Both lasers have a short-term laser line width
of several tens of kHz.  Their beams have a $1/e^2$ intensity
waist radius of 130\,$\mu$m at the molecular sample, through which
they propagate nearly collinearly. They are polarized parallel to
the magnetic bias field $B_z$ (pointing in the vertical direction)
and thus can only induce $\pi$ transitions. The frequencies of
both lasers are automatically read out using a commercial
wavemeter (WS7 from HighFinesse). The binding energy $E_B$ of
$\qmv$ minus the binding energy of our initial Feshbach state
(4.4\,MHz$\times h$) corresponds to the frequency difference of
the two lasers. The wavemeter has a nominal accuracy of 60\,MHz
after calibration. Over the course of days we have observed drifts
of $\pm200$\,MHz, e.g. by repeatedly addressing the same
spectroscopic line. Over the length of a few experimental cycles
(5\,minutes) the wavemeter is stable to within 10\,MHz, which
represents a random noise floor. Assuming a sufficiently smooth
behavior of the wavemeter, drifts of the wavemeter typically
affect the frequency measurements of laser 1 and laser 2 in a
similar way, especially since the laser frequencies only differ by
3\%. Thus, such common mode drifts cancel out in the binding
energy to first approximation. Indeed, based on our experience
with the wavemeter where we have measured binding energies of a
few sharp lines over an extended period of time and via various
intermediate levels we estimate the accuracy to reach 30\,MHz.
This includes peak position uncertainties due to variations of the
number of molecules produced, as well as a frequency drift of
laser 2 during the time between the laser pulse and wavelength
measurement.

\begin{figure}[tbp] \centering \includegraphics{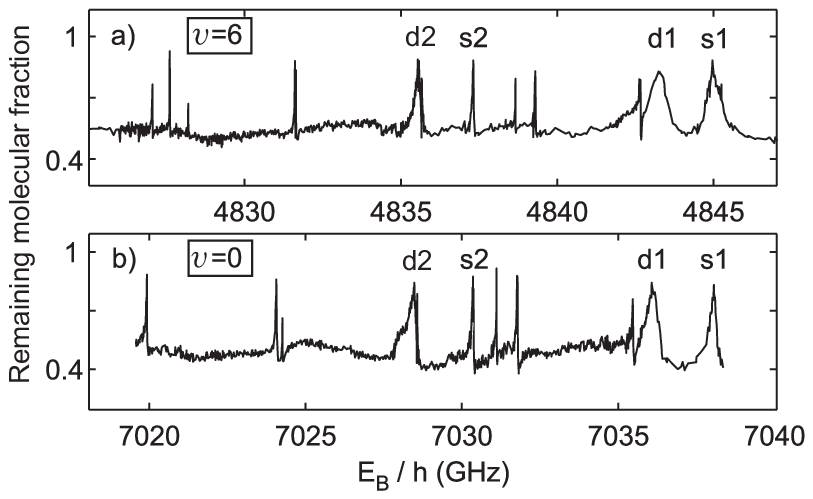}
\caption{Scans of two vibrational levels: a)  $\vib = 6$, b)
$\vib = 0$. Plotted is the remaining molecular fraction as a
function of the binding energy $E_B$, which basically corresponds
to the laser difference frequency. The scans were recorded using
an excited state $\qme$ with $\clg$ character. The labels s and d
indicate rotational states $N=0$ and 2, respectively. The numbers
after the labels s or d indicate the position in the spectrum.}
\label{fig:v0_v6_1g}
\end{figure}

Fig.\,\ref{fig:vibladder} a) shows the measured binding energies
of the triplet potential as a function of the vibrational quantum
number $\vib$ at a magnetic field of 1005.8\,G. There are 41
vibrational states with binding energies ranging from
5\,MHz$\times h$ to about 7038\,GHz$\times h$.  The vibrational
splitting between the two lowest vibrational levels, $\vib=0$ and
$\vib=1$, is about 393\,GHz. Each vibrational state has hyperfine,
rotational, and Zeeman substructure. This structure is spread out
over a range of about 20\,GHz as shown in Fig.\,\ref{fig:v0_v6_1g}
for the states $\vib=0$ and $6$. These spectra typically consist
of roughly 1000 points corresponding to an average step size of 20
MHz. Each point represents one production and measurement cycle
which takes 28\,s.

In each spectrum of Fig.\,\ref{fig:v0_v6_1g} we observe some 10
lines which vary markedly in line width.  The width of each line
is determined by the coupling between the levels $\qme$ and
$\qmv$, \emph{i.e.} the Rabi frequency $\Omega_2$. Interestingly,
for our measuring scheme this width scales as $\Omega_2^2$
\cite{DissLang09} and not as $\Omega_2$ \cite{width, Sho90}, as
one might expect. We have taken advantage of this enhanced
broadening when searching for lines and vibrational manifolds,
which otherwise can be like looking for a needle in a haystack.
For example, for $\vib=0$ and an intensity $I_2$ of a few hundred
mW we reached line widths of several GHz. The substructure of the
desired vibrational level then appears essentially as a single
broad line with a width of about 20 GHz. Once this level was found
the power was reduced, in order to resolve its substructure.

In general, we expect spectra of different vibrational levels
to be very similar. Up to $\vib=35$ this is indeed the case. For
$\vib \ge 35$ the vibrational manifolds start to overlap, as the
splitting between them becomes smaller than 20 GHz. Spectra a)
($\vib=0$) and b) ($\vib=6$) of Fig.\,\ref{fig:v0_v6_1g} are
clearly similar. Some visible differences are artifacts, e.g. it
seems that in spectrum b) two narrow lines are missing at $E_B / h
\approx$\,7021GHz. This can be explained by the fact that the
lines were narrower (width $\leq10$\,MHz) than the local step size
of that scan and thus escaped observation.

\section{Quantum numbers and assignment}
\label{quantum}
\subsection{Quantum numbers}
\label{numbers}

For a deeper understanding of the structure of the
spectrum and its assignment we now discuss the relevant quantum
numbers and selection rules.

For the weakly bound levels like the Feshbach state or the
vibrational levels close to the atomic asymptote Hund's coupling
case (e) with atomic quantum numbers is most appropriate. Here
none of the angular momenta couple to the molecular axis. The
state vector is described by
\begin{align*}
|(f_A,f_B)f, l, F, M\rangle\,,
\end{align*}
\noindent where $f_A,f_B$ are the total angular momenta for atoms
A and B, $f$ is the sum of both atomic angular momenta, $l$ is the
mechanical rotation of the atomic pair, $F$ is the total angular
momentum of the pair and $M$ is its projection onto a space fixed
axis. At low magnetic fields the prepared Feshbach state $\qmf$
can be approximated by the state vector $|f_A=2, f_B=2, f = 2,
l=0, F=2, M=2\rangle$. At the magnetic field used in the
experiment (1005.8\,G), $F$ is no longer a good quantum number
because of only weak coupling between $l$ and $f$. A more
appropriate state vector is
\begin{align*}
|(f_A,f_B)f, m_f, l, m_l, M\rangle.
\end{align*}
$M$ is a good quantum number, while for example, $f_A$ and $f_B$
have expectation values \cite{expectval} of about 1.79 instead of
quantum numbers $f_A=f_B=2$ for 0\,G.

Due to the strong hyperfine coupling of Rb and the large exchange
energy, deeply bound levels of the triplet state can be described
by Hund's coupling case (b$_\beta$) at low magnetic fields
\cite{Tow55}, namely:
\begin{align*}
|N, (I,S)f, F, M\rangle\,,
\end{align*}
\noindent where $I$ and $S$ are the total nuclear and electronic
spin quantum numbers, $N$ the molecular rotation including
electron orbital angular momentum, and $F$ and $M$ have exactly
the same meaning as in Hund's case (e). The index $\beta$
indicates that the nuclear spin I is not coupled to the molecular
axis. Since both atoms are in the 5$S$ configuration of Rb, the
molecular electronic orbital angular momentum is zero. This means
that $N=l$ and that $f$ is the same as in Hund's case (e). Here we
get
\begin{align*}
|N, m_N, (I,S)f, m_f, M\rangle\,,
\end{align*}
as an appropriate state vector for higher magnetic fields, where
$M=m_N+m_f$. One can show that owing to the antisymmetry of the
molecular wave function with respect to nuclear exchange (nuclear
spin of $^{87}$Rb $i=\frac{3}{2}$), molecules with even (odd) $N$
in the $\triplet$ state must have either a total nuclear spin
$I=1$ or $3$ ($I=0$ or $2$) \cite{Tow55, He50}. For the $\singlet$
singlet ground state this relation is reversed because it has $g$
symmetry in contrast to the $u$ symmetry of the triplet state. For
large magnetic fields, $f$ loses its meaning as $S$ and $I$ start
to decouple.

Expectation values for the total nuclear spin and the electron
spin  for our Feshbach level are  $I = 1.56$ and $S = 0.76$,
respectively. Thus we have significant electronic singlet-triplet
mixing and consequently also mixing of the basis vectors with
different $I$. In contrast, the excited intermediate state $\qme$
has well defined quantum numbers $S$ and $I$. Thus $\qme$ largely
determines which quantum numbers the deeply bound $\triplet$
levels will have in the Raman transition.

The intermediate level $\qme$ is located in the $\vib'=13$
manifold of the $\tripletex (5S_{1/2}+5P_{1/2})$ potential. Due to
significant effective spin-spin interaction, the vibrational
manifold is split into two components, $\clg$ and $\cog$,
separated by 47\,GHz. As intermediate level $\qme$ we either
choose a level with $\clg$ character or with $\cog$ character. Its
rotational energy must be low because the Feshbach level has the
lowest rotational quantum number $N=0$. As rotation is low in
$\qme$, decoupling of $S$ and $I$ from the molecular axis by
rotation is not yet important. We can then approximate the state
$\qme$ by quantum numbers for Hund's case (a$_\alpha$) where the
index $\alpha$ indicates that the nuclear spin $I$ is coupled to
the molecular axis,
\begin{align*}
|S\Sigma,I\Omega_I,F,M \rangle\,.
\end{align*}
\noindent The projections of the electronic and nuclear spin onto
the molecular axis appear as quantum numbers $\Sigma$ and
$\Omega_I$, respectively.

The $\qme$ level  with $\clg$ character is energetically the
lowest within the $\clg(v = 13)$ manifold. It has the quantum
numbers $S=1$, $\left|\Sigma\right|=1$, $I=3$,
$\left|\Omega_I\right| \approx3, $ $F \approx 2$ and $M=2$ for low
magnetic fields $B_z$. Its excitation energy is 294.6264(2)
THz$\times h$ with respect to $\qmf$ at a field of 1005.8G.

The  $\qme$ level with $\cog$ character has an excitation energy
of 294.6736(2)\,THz. It can be described by the quantum numbers
$S=1$, $\Sigma=0$, $I=3$, $M = 2$  and $F = 3$ at $B_z$ = 0.
Because of its low hyperfine coupling the better choice of basis
vector here is
\begin{align*}
|S\Sigma,(JI)F,M \rangle\,,
\end{align*}
where $J$ results from the coupling of the electronic spin and the
molecular rotation ($\mathbf{J}=\mathbf{N}+\mathbf{S}$). For our
$\cog$ $\qme$ level J is approximately zero. Our $\clg$ $\qme$
state is a superposition of N=1 and 3.

As stated before, due to the laser polarization along the magnetic
field only $\pi$ transitions are allowed, which results in the
selection rule $\Delta M = 0$. Further, in a one-photon transition
parity has to change. The Feshbach state $\qmf$ has a total parity
``plus"  because it is a $\Sigma^+$ state and $(-1)^N = 1$. Thus
we can only address $\qme$ levels with ''minus" parity and $\qmv$
levels with ''plus" parity. This means that the $\qmv$ level must
have an even rotational quantum number $N=0,\,2,\,4,\, ...$. For
the $\qme$ level only quantum numbers $N=1,\,3, ...$ or
superpositions of these are available. In fact, as the
$\tripletex$ state is well described in a Hund's case (a) basis,
$N$ is in general not a good quantum number for the $\qme$ level.
The selection rule $\Delta N = \pm1$ for $N$ determines the range
of reachable levels for $\qmv$ according to the superposition in
$\qme$. The selection rules $\Delta I = 0$ and $\Delta S = 0$ are
important for the transition $\qme$ to $\qmv$. For the transition
$\qmf$ to $\qme$ they are, however, nearly irrelevant since $I$
and $S$ are not good quantum numbers for level $\qmf$.

\subsection{Assignment of spectral lines}
\label{assign}

\begin{figure}[tbp] \centering \includegraphics{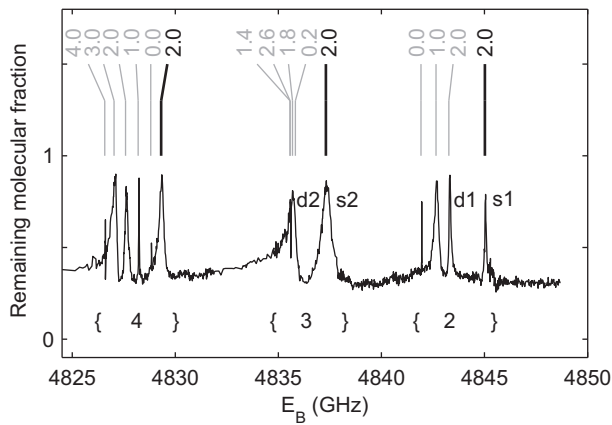}
\caption{Scan of the hyperfine, rotational and Zeeman structure in
the $\triplet(\vib=6)$ manifold using an excited level $\qme$ with
$\cog$-character. The x-axis shows the binding energy $E_B$. The
quantum number $f$ is shown below each of the three groups of
lines. Thick black lines above the spectrum indicate states with
$N=0$, grey lines correspond to states with $N=2$. The upper row
of numbers indicates the expectation values of the magnetic
quantum number $m_f$ at the magnetic field of 1005.8 G.}
\label{fig:v60g}
\end{figure}

Figure \ref{fig:v60g} shows measured and calculated (based on the
coupled-channel model, see section \ref{ccmodel}) lines of the
$\vib = 6$ spectrum, where the excited state $\qme$ with $0_g^-$
character was selected. The lines form three groups with the
quantum numbers $f= 4, 3, \text{and}~2$,  according to the
hyperfine coupling of $S=1$ and $I=3$. This is a clear indication
that the hyperfine energy is still dominant compared to the Zeeman
energy.

Each group consists of one line with $N = 0$  corresponding to a
non-rotating molecule, and several lines with  $N=2$ which are
shifted to lower binding energy by about 2 GHz$\times h$ due to
rotation. The fact that we do not observe lines with $N > 2$ can
be explained as follows: The excited state $\qme$ has ''minus"
total parity. Since $J = 0$ and $S = 1$, $N$ must be equal to one
and is also a good quantum number for this lowest level in the
$0_g^-$ manifold. Thus when using the state with $\cog$ character
only final states $\qmv$ with $N=0,\,2$ can be addressed in the
$\triplet$ state due to the selection rule $\Delta N = \pm1$.

The overall structure of each vibrational level can be
understood with a relatively simple effective Hamiltonian
\cite{Du72}:
\begin{align*}
H= \frac{A}{2}\mathbf{S}\cdot\mathbf{I}+B_v\mathbf{N}^2 + \mu_B
\,g_S \, S_z \, B_z
\end{align*}
\noindent with the atomic hyperfine structure constant $A = 3.42$
GHz$\times 4\pi^2 / h$ \cite{Biz99}, the total electronic and
nuclear spin operator $\mathbf{S}$ and $\mathbf{I}$, respectively,
and the operator for molecular rotation $\mathbf{N}$. $B_v$ is the
rotational constant of the desired vibrational level $v$. The last
term describes the Zeeman effect of the electronic spin when
exposed to an external magnetic field $B_z$ in z-direction. The
nuclear Zeeman term is neglected. The Zeeman effect can be
evaluated for the case of strong hyperfine coupling such that $f$
(with $\mathbf{f}=\mathbf{S}+\mathbf{I}$) is still a good quantum
number. Using a simple vector model for $\mathbf{f}$
 one can derive a Land\'e
factor $g_f$:
\begin{eqnarray}
H&= &\frac{A}{2}\mathbf{S}\cdot\mathbf{I}+B_v\mathbf{N}^2+\mu_B
\,g_f \, m_f \, B_z\nonumber\\
\text{with}&g_f&=\frac{g_S}{2}\left(1+\frac{S(S+1)-I(I+1)}{f(f+1)}\right)
\label{hfssimple}
\end{eqnarray}

For each $f$, the $N=2$ group is shifted to lower binding energy
by $6B_v$ compared to $N=0$, corresponding to about 2 GHz for the
$v = 6$ manifold (see Figure \ref{fig:v60g}). Each  $N=2$ group is
split by the Zeeman energy according to the $m_f$ quantum number
of each line. Only those lines which have quantum numbers where
$M=2=m_f+m_N$ can be observed. The level $f=3$ has a small $g_f $
factor (1/6 for $I = 3$), which gives rise to a small Zeeman
splitting. Additionally, one sees that the splitting between $f=4$
and $f=2$, which would be about 12 GHz for the pure hyperfine
part, is enlarged by a Zeeman contribution because of the
different signs of $g_f$ for $f=4$ and $f=2$. In Figure
\ref{fig:v60g} the expectation values for $m_f$ resulting from the
coupled channel model are given in the upper row and show that
these lead to good quantum numbers for $f=4$ and $f=2$ but are
less good for $f=3$, where the hyperfine and electronic Zeeman
energies are small. The mixing of states with different $m_f$ for
$f = 3$ is not included in the simple model of Eq.
\ref{hfssimple}.

As discussed before, for the intermediate state with $\cog$
character we exclusively observed $N = 0$ and $N = 2$ levels in
the $\triplet$ potential. This restriction does not necessarily
apply when using the intermediate level $\qme$ with $\clg$
character. The corresponding spectrum in \mbox{Figure
\ref{fig:v0_v6_1g} a)} shows the same $v = 6$ manifold as in
Figure \ref{fig:v60g}, only $\qme$ has $\clg$ instead of $\cog$
character. It has additional lines (at $\sim$4838\,GHz and at
$\sim$4832\,GHz) which match the predicted positions of $N = 4$
levels.

As mentioned in Section \ref{numbers} the level $\qme$ with $\clg$
character would be described by a superposition of $N=1$ and $N=3$
states in a Hund's case (b) basis. We thus expect to see
transitions to levels with $N=0, 2, 4$ of the $\triplet$ state
according to the selection rule $\Delta N=\pm 1$. In order to
avoid confusion, we note that the $N=0$ and $N=2$ lines between
4825 GHz and 4830 GHz which are clearly visible in Figure
\ref{fig:v60g} are weak or are not even observed in Figure
\ref{fig:v0_v6_1g}. For the same reasons as discussed at the end
of section \ref{exp}, the experimental step size of about 10 MHz
might have been too large compared to the narrow line widths for
these transitions to be seen.

It was not always necessary to carry out a complete scan as in
Figure \ref{fig:v60g} in order to assign quantum numbers to
observed lines in arbitrary vibrational levels. Often it was
sufficient to measure a few characteristic lines and splittings
and to compare them to the calculated spectrum. These data were
then used to optimize the coupled channel model along with its
$\triplet$ and $\singlet$ Born-Oppenheimer potentials.

\section{Coupled-channel model and optimization of the $\triplet$ potential}
\label{ccmodel}

In our work we use a coupled-channel model \cite{Du95, Tie98,
Kr90} that can calculate all bound states of the $\singlet$ and
$\triplet$ states, which correlate with the atomic asymptote
$5^2S_{1/2}+5^2S_{1/2}$. The program has helped us with the search
for lines as well as with their identification. Using our data we
are able to optimize the Born-Oppenheimer potential of the
$\triplet$ state as well as to improve the potential of the ground
state $\singlet$ given in \cite{seto:2000}.  In the following we
briefly describe the model and explain how the $\singlet $ and
$\triplet$ Born-Oppenheimer potentials are optimized.

To cover the full range of experimental data by a single
theoretical model, only a coupled channel analysis is adequate. It
includes the calculation of the molecular bound states as well as
the scattering resonances. It takes into account the $\singlet$ and
$\triplet$ potential functions, the hyperfine coupling, the Zeeman
interaction, rotation and the effective spin-spin interaction.
Such a theoretical approach is described in several papers, e.g.
\cite{pas07}.

For the present analysis we include our measurements covering 135
lines of the $\triplet$ state as well as data from other work. For
the $\singlet$ ground state we added data from Fourier transform
spectroscopy by Seto {\it et al.} \cite{seto:2000} with more than
12000 lines. We also include measurements of Feshbach resonances
for the three isotopologues $^{85}$Rb$_2$, $^{87}$Rb$_2$, and
$^{85}$Rb$^{87}$Rb \cite{marte:2002, roberts:2001, erhard:2003,
papp:2006}, the four asymptotic levels from \cite{wynar}, and
measurements from Fourier transform spectroscopy for the
$\triplet$ state reported by Beser {\it et al.} \cite{Be09}.

The full Hamiltonian (cf. \cite{mie00,lau02,pas07}) for a pair of
atoms $A$ and $B$ can be written in the form
\begin{align}
\label{eq:ham}
H &= T_n+U_{\rm X}(R)P_{\rm X} + U_{\rm a}(R)(1-P_{\rm X})\nonumber\\
&+a_A(R)\vec{s}_A\cdot\vec{i}_A+a_B(R)\vec{s}_B\cdot\vec{i}_B\nonumber\\
&+\left[(g_{sA}s_{zA}-g_{iA}i_{zA})+(g_{sB}s_{zB}-g_{iB}i_{zB})\right]
\mu_BB_z\nonumber\\ &+\frac{2}{3} \lambda(R)(3S_Z^2-S^2).
\end{align}
In the present case of the homonuclear molecule $^{87}$Rb$_2$ the
parameters with index A are equal to those of index B. The first
term in the first line shows the kinetic energy $T_n$ where we
take the atomic masses from recent tables by G. Audi {\it et al.}
\cite{aud03}. The next two terms describe the potential energies
$U_{\rm X}$ and $U_{\rm a}$ for the motion of the atoms, where
$P_{\rm X}$ and $1-P_{\rm X}$ are projection operators on to the
uncoupled states X and a, respectively. $R$ corresponds to the
internuclear separation of the two atoms. The second line shows
the hyperfine interaction between the atomic electron spins
$\vec{s}_{A,B}$ and the atomic nuclear spins $\vec{i}_{A,B}$. The
main contribution to the functions $a_{A,B}(R)$ is the Fermi
contact term. The $R$ dependence of the hyperfine parameters
accounts for several effects: It takes into account the electronic
distortions of one atom by the other, \emph{i.e.} the binding, and
an effective coupling of the electron spin of one atom with the
nuclear spin of the other atom. We start with $R$-independent
atomic coupling constants taken from a compilation by Arimondo
{\it et al.} \cite{ari77}. These constants are later refined by a
simple Ansatz for the $R$ dependence (discussed later in Eq.
\ref{eq:hfs}). We neglect the nuclear quadrupole moment in the
hyperfine interaction, which might come into play for deeply bound
levels. The third line in Eq. \ref{eq:ham} gives the Zeeman energy
from the coupling of the electron spin and the nuclear spin with
an external homogeneous magnetic field $B_z$ in $z$ direction. The
electronic and nuclear $g$-factors for the atomic ground state of
the Rb isotopes are taken from the report in \cite{ari77}. This
term couples states with different $f$ quantum numbers. The last
line contains the spin-spin interaction which couples different
$N$ states of basis (b). It is formed by the total molecular spin
$S$ and its projection on the molecule fixed axis $Z$. The
parameter $\lambda$ is a function of $R$, of which one part has
$1/R^3$ dependence as a result from the magnetic dipole-dipole
interaction. In addition, $\lambda$ contains contributions from
second order spin-orbit interactions. The final analysis showed
that such a contribution is significant within the achieved
experimental accuracy. For example, this was important for the
precise location of Feshbach resonances involving $l=1$ and $l=2$
bound levels.

The functional form of the two Born-Oppenheimer potentials,
$\singlet$ and $\triplet$, is split into three regions on the
internuclear separation axis $R$: the short range repulsive wall
($R < R_\mathrm{SR}$), the asymptotic long range region ($R >
R_\mathrm{LR}$), and the intermediate deeply bound region in
between. The analytic form of the potentials in the intermediate
range, $U_{IR}$, is described by a finite power expansion of a
nonlinear function $\xi$ which depends on the internuclear
separation $R$,
\begin{equation}
\label{eq:uanal} U_{\mathrm {IR}}(R)=\sum_{i=0}^{n}a_i\,\xi^i(R),
\end{equation}
\begin{equation}
\label{eq:xv} \xi(R)=\frac{R - R_m}{R + b\,R_m}.
\end{equation}
Here the $a_i$ are fitting parameters (see table \ref{pota}). We
choose $b$ and $R_m$ such that only few parameters $a_i$ are
needed for describing the steep slope at the short internuclear
separation side and the much smaller slope at the large $R$ side.
$R_m$ is chosen close to the value of the equilibrium separation.
The potential is extrapolated for $R < R_{SR}$  by the short range
part $U_{SR}$ with:
\begin{equation}
\label{eq:rep}
  U_{SR}(R)= u_1 + u_2/R^{N_s}.
\end{equation}
We adjust the parameters $u_1$ and $u_2$ to get a continuous
transition at $R_{\rm SR}$. The final fit uses $N_s\approx 4.5$
for both the  $\singlet$\ and $\triplet$ states.

For large internuclear distances ($R > R_{\rm LR}$) we adopted the
standard long range form of molecular potentials
\begin{equation}
\label{eq:lrexp} U_{\mathrm {LR}}(R)=-C_6/R^6-C_8/R^8
-C_{10}/R^{10}\pm E_{\mathrm{exch}},
\end{equation}
\noindent where the exchange contribution given by \cite{exc}
\begin{equation}
\label{eq:exch} E_{\mathrm{exch}}=A_{\mathrm{ex}} R^\gamma
\exp(-\beta R)
\end{equation}
is negative for the singlet and positive for the triplet
potential. By adjusting the parameter $a_0$ in Eq. \ref{eq:uanal}
we can  assure a continuous transition from $U_{\mathrm {LR}}$ to
$U_{\mathrm {IR}}$. As mentioned, the data on hand include three
different isotopologues, namely $^{85}$Rb$_2$, $^{87}$Rb$_2$, and
$^{85}$Rb$^{87}$Rb. Using a model developed earlier for LiK
\cite{tie:2009} we checked in the final calculations  that the
data are not sufficiently precise to extract deviations from the
Born-Oppenheimer approximation. Thus we derive the potentials
without any mass scaling correction.

Having discussed the complete physical model, we are now ready to
calculate all the relevant bound state energies and scattering
properties to compare them with the experimental data. We have
decided to evaluate the Hamiltonian in Hund's basis (e) as $f$ is
still a relatively good quantum number. This is due to the fact,
that the hyperfine interaction is larger than the Zeeman
interaction in our experiment. As the total electron spin is not a
good quantum number in Hund's case (e), the choice of this basis
leads to significant non-diagonal matrix elements from the
Born-Oppenheimer potentials, which are given for a pure singlet or
triplet state.

We evaluate the free parameters of the model with a
self-consistent iteration loop and alternate between: (i) coupled
channel calculations of the full Hamiltonian and (ii) solving the
Schr\"odinger equation separately for the states $\singlet$ and
$\triplet$ using only the first line of Eq.\,\ref{eq:ham} and
applying the Numerov procedure. The coupled channel calculations
in step (i) are used to determine the hyperfine and Zeeman
structure. From this we construct hyperfine free spectroscopic
data. This data is the input for the Born-Oppenheimer potential
fits in step (ii). We incorporate the fitting routine for the
plain Born-Oppenheimer potentials in step (ii). The optimization
of the singlet and triplet potentials is done simultaneously as
both potentials have a common asymptote. These asymptotic
potentials are given in Eq.\,\ref{eq:lrexp} with equal dispersion
coefficients.

Normally, the iteration loop between the potential function fit
and the coupled channel calculation for producing hyperfine
structure free data converges in a few steps, but we observed some
systematic deviations between measurements and calculations.
Specifically the hyperfine splitting showed variations of a few
percent within the vibrational ladder. Even though our
experimental accuracy is not better than 30\,MHz such small
variations are observable since the hyperfine splitting in
$^{87}$Rb$_2$ of 12\,GHz is so large.

In order to theoretically account for the hyperfine variations we
extended the model. We changed the fixed atomic hyperfine
parameters to a function in $R$. We chose a function which
switches at a distance $R_0$ from the atomic value of the
hyperfine constant to another value for a deeply bound dimer,
\begin{equation}
\label{eq:hfs}
a_{A/B}=a_{Rb}\left(1+ \frac{c_f}{e^{(R-R_0)/\Delta R} +1}\right).
\end{equation}
Here $a_{Rb}$ is the atomic hyperfine constant, $c_f$ the
fractional change of the constant and $R_0$ and $\Delta R$
describe the switching distance and its width, respectively. We
also tried several other simple switching functions, which
produced about equal fit quality. The function in
Eq.\,\ref{eq:hfs} is easily applicable for other isotopes by
introducing the proper atomic hyperfine constant, because the
scaling $c_f$ will be independent of the isotope. We chose
$R_0=11.0~a_0$ and $\Delta R=0.5~a_0$ (where $a_0=0.5292 \times
10^{-10}$\,m is the Bohr radius), such that switching takes place
approximately at the minimum of the $\triplet$ potential. From our
fits we obtained an amplitude $c_f= -0.0778$, which corresponds to
a variation of the hyperfine coupling across the potential depth
of up to 8\%. The influence of this hyperfine variation within an
individual vibrational manifold will be smaller because of the
averaging of Eq.\,\ref{eq:hfs} over the vibrational wave function.
As mentioned before, also the spin-spin interaction needs
optimization in order to explain systematic shifts of Feshbach
resonances in s-wave and p-wave scattering channels, as measured
in \cite{marte:2002}. The spin-spin interaction couples different
 partial waves $l$ subject to the selection rule $\Delta
l=0,\,\pm2$ such that resonances in a s-wave scattering channel
involve bound states with $l=0$ and $l=2$. The spin-spin
interaction splits the resonances according to $\left|m_l\right|$,
the projection quantum number of rotation on the space fixed axis.
We use a simple functional form for $\lambda(R)$ in Eq.
\ref{eq:ham}
\begin{equation}
\label{lambda} \lambda(R) = -\frac{3}{4}
\alpha^2\left(\frac{1}{R^3}+a_\mathrm{SO}
\exp{\left(-b(R-R_\mathrm{SO})\right)}\right),
\end{equation}
%made consistent.  R dep.
that consists of two terms. The first term represents the magnetic
dipole-dipole interaction and the second term is the second order
spin-orbit contribution. If Eq.\,\ref{lambda} is given in atomic
units, $\alpha$ is the fine structure constant. Since the few data
at hand cannot be highly sensitive to the actual function, we
adopted values for $b$ and $R_\mathrm{SO}$ from a theoretical
approach by Mies {\it et al.} for Rb$_2$ \cite{Mie:96} (b= 0.7196
a$_0^{-1}$ and R$_\mathrm{SO}$= 7.5 a$_0$). We fitted the
parameter $a_\mathrm {SO}$, which yielded $a_\mathrm{SO}$= -0.0416
a$_0^{-3}$. With these parameters the second part in Eq.
\ref{lambda} contributes significantly to the effective spin-spin
interaction in the internuclear separation interval $R < 20 a_0 $.
This affects bound levels of the triplet state that determine the
Feshbach resonances. After optimization we achieve an accuracy of
0.1\,G for the position of the Feshbach resonance. This correction
to the constant $\lambda$ does not influence the description of
the other bound states within their uncertainty.

Finally, we found that we can improve the fit by adding to the
long range formula Eq.\,6 a term of the form $-C_{26}/R^{26}$, and
retaining all constraints such as continuity as before. From our
fits we obtain an amplitude $C_{26}$ which contributes about a
thousandth of the total long range energy at the connection point
$R_{LR}$. The exponent of 26 was chosen so that the term is
negligible outside a small region around the long range connection
point $R=R_{LR}$.

The final parameter sets of the potentials are shown in table
\ref{potX} and \ref{pota} for the singlet and triplet states,
respectively. The derived potentials and the corrections defined
above agree very closely with all observations to within their
experimental uncertainties, and the normalized standard deviation
(\emph{i.e.} standard deviation / experimental uncertainty) is
close to one. In fact, depending on which of our experimental data
sets (hyperfine free spectra, binding energies from dark state
spectroscopy or Feshbach resonances) we compare to the model
calculations, the normalized standard deviations only vary
slightly, ranging from 1.01 to 1.3. The normalized standard
deviation from the joint calculation over the huge body of
ro-vibrational energies (12459 data points from the states
$\singlet$ and $\triplet$) is about 1.01, which is quite
satisfactory.

As another result of our analysis we were able to eliminate an
ambiguity in  the rotational assignment of the Fourier transform
spectroscopy data reported by Beser {\it et al.} for the state
$\triplet$ \cite{Be09}. The authors stated in the
rotational assignment an ambiguity by $\Delta N=\pm1$. We determined
that the shift of the rotational quantum number must be
$\Delta N=+1$. Afterwards we used the data from \cite{Be09} with
that assignment for the further fits. The result in Table
\ref{pota} includes these data and will give a better confidence
in predicting rotational levels with higher rotational quantum
numbers.

\begin{table}
\fontsize{8pt}{13pt}\selectfont \caption{Parameters of the
analytic representation of the potential of state $\singlet$. The
energy reference is the dissociation asymptote and the term energy
$T_e^X$ is the depth of the potential. Parameters with $\ast$ are
set for continuous extrapolation of the potential and those with
$\ast\ast$ see text for interpretation.}
\label{potX}   % pot used X41
\begin{tabular*}{1.0\columnwidth}{@{\extracolsep{\fill}}|lr|}
\hline
   \multicolumn{2}{|c|}{$R < R_\mathrm{SR}=$ 3.126 \AA}    \\
\hline
   $N_s$      & 4.53389\\
   $u_1^\ast$ & -0.638904880$\times 10^{4}$ \wn \\
   $u_2^\ast$ & 0.112005361$\times 10^{7}$  \wn \AA $^{N_s}$ \\
\hline
   \multicolumn{2}{|c|}{$R_\mathrm{SR} \leq R \leq R_\mathrm{LR}=$ 11.000 \AA}    \\
\hline
    $b$ &   $-0.13$              \\
    $R_\mathrm{m}$ & 4.209912760 \AA               \\
    $a_{0}$ &  -3993.592873 \wn\\
    $a_{1}$ &  0.000000000000000000$ $ \wn\\
    $a_{2}$ &  0.282069372972346137$\times 10^{ 5}$ \wn\\
    $a_{3}$ &  0.560425000209256905$\times 10^{ 4}$ \wn\\
    $a_{4}$ & -0.423962138510562945$\times 10^{ 5}$ \wn\\
    $a_{5}$ & -0.598558066508841584$\times 10^{ 5}$ \wn\\
    $a_{6}$ & -0.162613532034769596$\times 10^{ 5}$ \wn\\
    $a_{7}$ & -0.405142102246254944$\times 10^{ 5}$ \wn\\
    $a_{8}$ &  0.195237415352729586$\times 10^{ 6}$ \wn\\
    $a_{9}$ &  0.413823663033582852$\times 10^{ 6}$ \wn\\
   $a_{10}$ & -0.425543284828921501$\times 10^{ 7}$ \wn\\
   $a_{11}$ &  0.546674790157210198$\times 10^{ 6}$ \wn\\
   $a_{12}$ &  0.663194778861331940$\times 10^{ 8}$ \wn\\
   $a_{13}$ & -0.558341849704095051$\times 10^{ 8}$ \wn\\
   $a_{14}$ & -0.573987344918535471$\times 10^{ 9}$ \wn\\
   $a_{15}$ &  0.102010964189156187$\times 10^{10}$ \wn\\
   $a_{16}$ &  0.300040150506311035$\times 10^{10}$ \wn\\
   $a_{17}$ & -0.893187252759830856$\times 10^{10}$ \wn\\
   $a_{18}$ & -0.736002541483347511$\times 10^{10}$ \wn\\
   $a_{19}$ &  0.423130460980355225$\times 10^{11}$ \wn\\
   $a_{20}$ & -0.786351477693491840$\times 10^{10}$ \wn\\
   $a_{21}$ & -0.102470557344862152$\times 10^{12}$ \wn\\
   $a_{22}$ &  0.895155811349267578$\times 10^{11}$ \wn\\
   $a_{23}$ &  0.830355322355692902$\times 10^{11}$ \wn\\
   $a_{24}$ & -0.150102297761234375$\times 10^{12}$ \wn\\
   $a_{25}$ &  0.586778574293387070$\times 10^{11}$ \wn\\
\hline
   \multicolumn{2}{|c|}{$ R > R_\mathrm{LR}$}\\
\hline
 % ${D_{asymptote}=U_\infty}$ & 0.0 \wn    \\
 ${C_6}$ &    0.2270032$\times 10^{8}$ \wn\AA$^6$      \\
 ${C_{8}}$ &  0.7782886$\times 10^{9}$ \wn\AA$^8$   \\
 ${C_{10}}$ & 0.2868869$\times 10^{11}$ \wn\AA$^{10}$   \\
 ${C_{26}^{\ast\ast}}$ & 0.2819810$\times 10^{26}$ \wn\AA$^{26}$   \\
 ${A_{ex}}$ & 0.1317786$\times 10^{5}$ \wn\AA$^{-\gamma}$   \\
 ${\gamma}$ & 5.317689    \\
 ${\beta}$ & 2.093816 \AA$^{-1}$   \\
\hline
 \multicolumn{2}{|c|}{Derived constants:} \\
\hline
\multicolumn{2}{|l|}{equilibrium distance:\hspace{2.2cm} $R_e^X$= 4.20991(5) \AA} \\
\multicolumn{2}{|l|}{electronic term energy:\hspace{1.6cm}     $T_e^X$= -3993.5928(30) \wn}\\
\hline
\end{tabular*}
\end{table}

\begin{table}
\fontsize{8pt}{13pt}\selectfont \caption{Parameters of the
analytic representation of the potential of state $\triplet$. The
energy reference is the dissociation asymptote and the term energy
$T_e^a$ is the depth of the potential. Parameters with $\ast$ are
set for continuous extrapolation of the potential and those with
$\ast\ast$ see text for interpretation.}
\label{pota}    % pot used a41
\begin{tabular*}{1.0\columnwidth}{@{\extracolsep{\fill}}|lr|}
\hline
   \multicolumn{2}{|c|}{$R < R_\mathrm{SR}=$ 5.07 \AA}    \\
\hline
   $N_s$      & 4.5338950\\
   $u_1^\ast$ & -0.619088543$\times 10^{3}$ \wn \\
   $u_2^\ast$ & 0.956231677$\times 10^{6}$  \wn \AA $^{N_s}$ \\
\hline
   \multicolumn{2}{|c|}{$R_\mathrm{SR} \leq R \leq R_\mathrm{LR}=$ 11.00 \AA}    \\
\hline
    $b$ &   $-0.33$              \\
    $R_\mathrm{m}$ & 6.0933451 \AA               \\
    $a_{0}$ &  -241.503352 \wn\\
    $a_{1}$ & -0.672503402304666542$ $ \wn\\
    $a_{2}$ &  0.195494577140503543$\times 10^{ 4}$ \wn\\
    $a_{3}$ & -0.141544168453406223$\times 10^{ 4}$ \wn\\
    $a_{4}$ & -0.221166468149940465$\times 10^{ 4}$ \wn\\
    $a_{5}$ &  0.165443726445793004$\times 10^{ 4}$ \wn\\
    $a_{6}$ & -0.596412188910614259$\times 10^{ 4}$ \wn\\
    $a_{7}$ &  0.654481694231538040$\times 10^{ 4}$ \wn\\
    $a_{8}$ &  0.261413416681972012$\times 10^{ 5}$ \wn\\
    $a_{9}$ & -0.349701859112702878$\times 10^{ 5}$ \wn\\
   $a_{10}$ & -0.328185277155018630$\times 10^{ 5}$ \wn\\
   $a_{11}$ &  0.790208849885562522$\times 10^{ 5}$ \wn\\
   $a_{12}$ & -0.398783520249289213$\times 10^{ 5}$ \wn\\
\hline
   \multicolumn{2}{|c|}{$ R > R_\mathrm{LR}$}\\
\hline
 % ${D_{asymptote}=U_\infty}$ & 0.0 \wn    \\
 ${C_6}$ &    0.2270032$\times 10^{8}$ \wn\AA$^6$      \\
 ${C_{8}}$ &  0.7782886$\times 10^{9}$ \wn\AA$^8$   \\
 ${C_{10}}$ & 0.2868869$\times 10^{11}$ \wn\AA$^{10}$   \\
 ${C_{26}^{\ast\ast}}$ & 0.2819810$\times 10^{26}$ \wn\AA$^{26}$   \\
 ${A_{ex}}$ &0.1317786$\times 10^{5}$ \wn\AA$^{-\gamma}$   \\
 ${\gamma}$ & 5.317689    \\
 ${\beta}$ & 2.093816 \AA$^{-1}$   \\
\hline
 \multicolumn{2}{|c|}{Derived constants:} \\
\hline
\multicolumn{2}{|l|}{equilibrium distance:\hspace{2.2cm} $R_e^a$= 6.0940(10) \AA} \\
\multicolumn{2}{|l|}{electronic term energy:\hspace{1.6cm}     $T_e^a$= -241.5034(30) \wn}\\
\hline
\end{tabular*}
\end{table}

\section{Progression of vibrational levels and their substructure}
\label{vibmode}

In the following we will discuss interesting insights which we
have gained from our analysis of the coupled system. In particular
we investigate the progression of several quantities in the
vibrational ladder and the mixing of singlet and triplet states.
We concentrate on the details of the $\triplet$ state, because it
was studied in full resolution of hyperfine and Zeeman energy over
the whole vibrational ladder for the first time. Such a body of
data does not exist for any other alkali dimer.

\subsection{Vibrational ladder and rotational progression}
\label{sec:vibladder} We return  to the vibrational states shown
in Fig.\,\ref{fig:vibladder}, which allows us to compare in detail
the optimized coupled-channel model with our experimental findings
for  all vibrational states.

\begin{figure}[htbp]
\centering \includegraphics{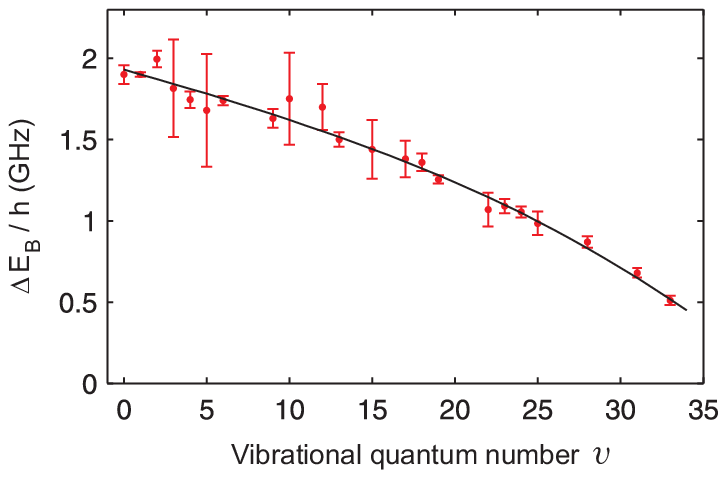} \caption{Energy splitting
$\Delta E_B$ between the ``s1"-level ($N=0$) and the ``d1"-level
($N=2$) for a given vibrational quantum number $v$. Large error
bars correspond to early measurements without simultaneous
wavemeter readings (see text). The continuous line is a
calculation based on the coupled channel model.}
\label{fig:vibladders1d1}
\end{figure}

The binding energies given in Fig.\,\ref{fig:vibladder} a)
correspond to the most deeply bound level in each vibrational
level, \emph{i.e.} the state with quantum numbers $N = 0$, $f = 2$
(at 0\,G) and $M = 2$. We will refer to this level as ``s1". The
``s1" level of $v=0$ in the $\triplet$ potential is also the
lowest bound state in that potential and has an observed binding
energy of $(7038.067\pm0.050)$\,GHz$\times h$ at 1005.8\,G with
respect to the lowest atomic asymptote $f_A\approx1, m_A=1$ and
$f_B\approx1, m_B=1$.  A list of all the other measured and
calculated bound state energies together with their expectation
values of the quantum numbers can be found in the supplementary
material \cite{EPAPS}.

Fig.\,\ref{fig:vibladder} b) shows the residues between our
experimental ``s1" data and the optimized model. In general the
model agrees very well with the measurements to within the error
bars. The data points with larger error bars belong to early
measurements without simultaneous wavemeter reading of both
lasers, which leads to a significant increase in the experimental
uncertainty (see section \ref{exp}).

We now investigate the progression of the rotational splitting in
the vibrational ladder. For this, we consider the ``s1" level
($N=0$) and its nearest neighbor in our spectra with ($N =2$)
which we call ``d1" (see Figs. \ref{fig:v0_v6_1g} and
\ref{fig:v60g}). The ``s1" and ``d1" levels both have $f=2$ and
$m_f = 2$. Thus the splitting between them is to a high degree
rotational energy. The splitting decreases with increasing $v$,
which is due to the fact, that the mean distance between the
Rb$_2$ nuclei (and hence the effective moment of inertia)
increases as $v$ increases. We have directly observed this
behavior in our experiments (see Fig. \,\ref{fig:vibladders1d1}).
There is very good agreement between the experimental data and the
calculation (continuous line) using the optimized potential of the
$\triplet$ state.

Besides the ``s1" level ($N=0, f = 2$) and the ``d1" level ($N=2,
f = 2$) we also observe states with $N=4$ and $f= 2$ for several
low lying vibrational levels. The two lines in Fig. 3 a) at 4839
GHz and the two lines in Fig. 3  b) at 7032 GHz have $N=4$.
Observation of these levels improved the precision when fixing the
position of the $\triplet$ potential minimum in terms of the
internuclear distance, or  the effective rotational constant
$B_v$. This turned out to be important for the reassignment of the
observations by Beser {\it et al.} \cite{Be09}, as discussed at
the end of section IV.

\subsection{ Hyperfine splitting and singlet-triplet mixing}
\label{hfsandmixing}

\begin{figure}[tbp]
\centering
\includegraphics{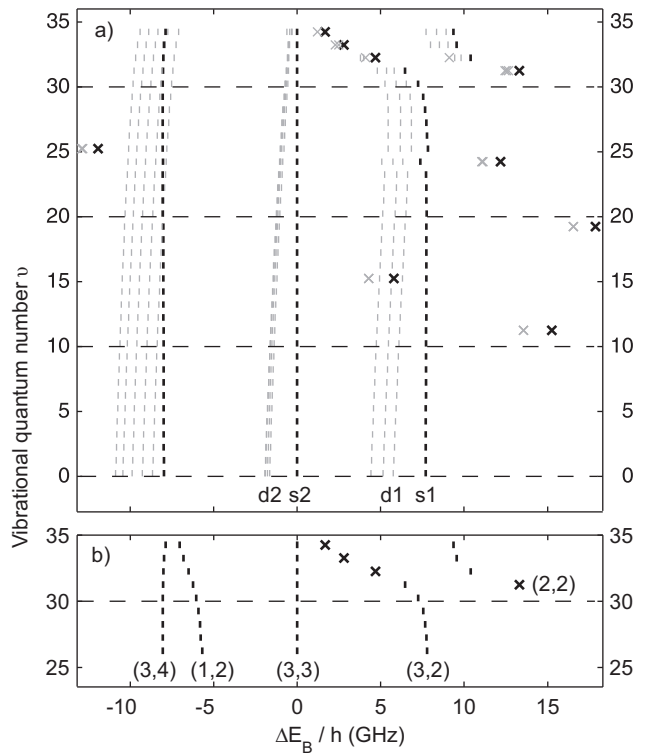}
\caption{a) Progression of the substructure of the vibrational
manifold at $B$ = 1005.8 G. Shown are calculated $\triplet$ levels
with quantum numbers $I \approx 3$, $M = 2$, $N = 0, 2$ and $v =
0...\,34$. Thick black lines correspond to $N=0$, grey ones to
$N=2$. The ``s2" level serves as energy reference ($\Delta E_B = 0
$) in each vibrational substructure. In addition to the triplet
levels nearby singlet levels of the $\singlet$ potential are also
shown (Thick black x's: $N=0$, grey x's: $N=2$). \\b) Symbols as
in part a). To discuss the singlet-triplet mixing from a
theoretical point of view we now also include lines with
$I\approx1$ but only $N=0$ for clarity. On the bottom, the
approximate quantum numbers $(I,f)$ are given. See text for
details.} \label{fig:alltheory}
\end{figure}

It is instructive to investigate the progression of the multiplet
structure of the vibrational levels. Figure
\ref{fig:alltheory}\,a) shows calculated levels for $M=2$ stacked
on top of each other for increasing $v$ of the $\triplet$
potential. We concentrate on the triplet states ($S = 1$) with
rotation $N=0,\,2$ and we restrict ourselves further to triplet
levels with $I = 3$. Additionally, we show singlet levels ($S=0$)
with $I = 2$ that are located in the vicinity of the triplet
levels. Thus, the typical stick spectrum of each vibrational
manifold in the $\triplet$ state looks like the one in Fig.
\ref{fig:v60g}. There are three groups of lines with $f=4,3,2$,
respectively. The vibrational quantum number $v$ runs from 0 to
34, where the multiplets do not yet overlap for different
vibrational levels.

In order to properly stack the data, we have chosen the level with
$N=0$, $f \approx 3$ to be the energy reference ($\Delta E_B = 0
$) for each vibrational level. We call this level ``s2" (see also
Figs. 3 and 4). This level is quite insensitive to mixing with
singlet levels which makes it a good reference because there is no
singlet level with $f=3$ and even parity for direct coupling. It
is clear from Fig. 6 a), that the multiplet structure of different
vibrational manifolds is similar, at least from $v$ = 0 to about
$v = 30$ as expected from the simple model Hamiltonian in
Eq.\,\ref{hfssimple}. The structure changes in a smooth and
monotonical way with $v$. For each of the quantum numbers
$f=4,\,3,\,2$ the splitting between $N=0$ and $N=2$ decreases with
increasing $v$ as discussed previously in
section\,\ref{sec:vibladder}.

Appreciable mixing of a singlet and a triplet level can occur when
the two levels with $f=2$ are located energetically closely
enough, in our case within a few GHz, corresponding to the
strength of the hyperfine interaction. One effect of mixing is a
shift of the level positions due to level repulsion, clearly seen
in Fig. \,\ref{fig:alltheory}. For $v=24$, for example, the
triplet lines of $f = 2$ are pushed to the left by about 0.4\,GHz
by the close singlet levels on the right hand side. For $v < 20$
we also observe narrow coincidences between triplet and singlet
levels, e.g. for $v=15$. Here almost no perturbation appears in
the graph, in contrast to the case $v=24$. We attribute this to a
significantly lower overlap of the vibrational wavefunctions of
the singlet and triplet levels for the case $v=15$ compared to
that of $v=24$. For vibrational quantum numbers $v > 30$ mixing is
very strong and happens for every vibrational level. Because the
long range behaviour of the $\triplet$ and $\singlet$ states is
similar, the overlap of the wave functions will become large for
high $v$.  The vibrational spacing will become similar to the
hyperfine splitting, and the state vectors here are best described
by Hund's coupling case (e) \emph{i.e.} quantum numbers of atom
pairs.

Singlet-triplet mixing not only occurs for triplet molecules with
$I = 3$ but also for $I = 1$. Figure \ref{fig:alltheory} b) shows
($N = 0$) triplet levels with $I = 3$ and $I = 1$ for $v = 25 $ to
34. The repulsion of the $I = 1$ levels from the singlet levels is
clearly visible. The figure shows an avoided-crossing-like
behavior for the levels on the right as a function of $v$,
indicating the strong mixing between $I=3$ and $I=2$ which also
means that u/g symmetry is broken for these levels. Further, our
calculations show that only levels with the same $f$ and $N$
quantum numbers mix considerably. The singlet lines shown here
have $f = 2$ and $N = 0$. Apparently, despite the relatively
strong magnetic fields of about 1000 G $f$ is still quite a good
quantum number. Indeed, $f$ appears as quantum number in both
state vectors for Hund's case (b) and (e) and loses its meaning
only for much higher magnetic fields. $N$ is good due to the small
effective spin-spin interaction.

\begin{figure}[tbp] \centering \includegraphics{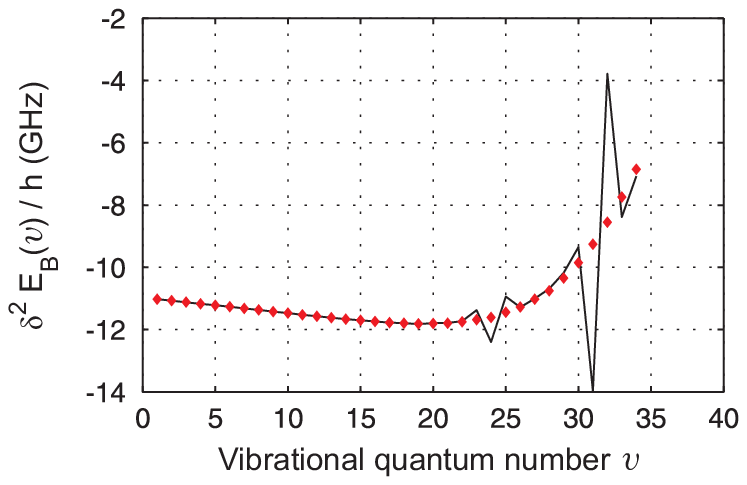}
\caption{ ``Discrete second derivative" of $E_B(\vib)$,
\emph{i.e.} $\delta^2 E_B(v) = E_B(v+1) - 2 E_B(v) + E_B(v-1) $.
The solid curve
 connects the points from the ``s1" levels while the  diamonds
correspond to the ``s2"-state.}
\label{fig:dev2_s1}
\end{figure}

The shift in position due to mixing can be traced more clearly
with a simple difference method discussed in the following.
Fig.\,\ref{fig:dev2_s1} shows the ``discrete second derivative"
$\delta^2 E_B(v)$ of the function for the binding energies
$E_B(v)$, \emph{i.e.} $\delta^2 E_B(v) = E_B(v+1) - 2 E_B(v) +
E_B(v-1)$. In other words it is the difference between the
neighboring energy splittings. The curve calculated from binding
energies ``s1" exhibits sudden jumps for particular vibrational
levels. These are due to singlet-triplet mixing and the $v$
positions are consistent with Fig.\,\ref{fig:alltheory}. In
contrast, the $ \delta^2 E_B(v)$ curve for the ``s2"-state is
smooth, and thus does not indicate mixing with the singlet lines,
which justifies its choice as energy reference in
Fig.\,\ref{fig:alltheory}.

\begin{figure}[tp]
\centering
\includegraphics{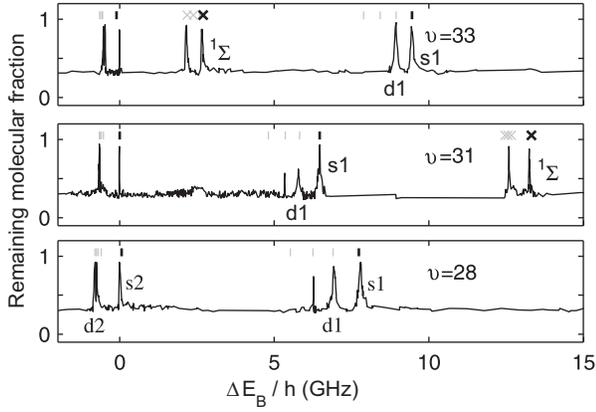}
\caption{Observation of strongly perturbed singlet levels. We
compare sections of line spectra for the vibrational levels $v =
28, 31$ and $33$. As in Fig. \ref{fig:alltheory} the ``s2"-level
is chosen as energy reference $\Delta E_B = 0 $.  The lines (and
crosses) above the experimentally observed spectrum are from
coupled-channel calculations. Black lines (crosses) represent
$N=0$ levels, while thin grey lines (crosses) correspond to $N=2$
levels for $S=1$ ($S=0$ respectively). The lines that originate
from the singlet states are labeled with `` $^1\Sigma$ ".}
\label{fig:v28to33_1g}
\end{figure}

We have confirmed the singlet-triplet mixing experimentally.
Fig.\,\ref{fig:v28to33_1g} shows scans of parts of the vibrational
levels $v = 28, 31, 33$ where the intermediate level $\qme$ with
$\clg$ character was used (see also Fig.\,\ref{fig:v0_v6_1g}). The
``s2"-level is chosen to be the energy reference at $\Delta E_B =
0 $ as before. For $v = 28$ we observe a line structure similar to
that of Figs. \ref{fig:v0_v6_1g} and \ref{fig:v60g}. Thus we take
the $v=28$ spectrum as reference of the pure case "triplet". In
fact, from our calculations we see that the next singlet level is
located about 30 GHz away. This detuning gives rise to only a very
small mixing which slightly lowers the triplet character of the
``s1"-level to 0.99. The singlet level obtains a triplet character
of 0.02. However, due to the selection rule $\Delta S = 0$, the
transition to the intermediate state $\qme$ with its very pure
triplet character would simply be too weak to see.

This situation changes drastically for $v = 31$ and $33$. Two
additional lines are visible in the spectrum which originate from
singlet states with $M = 2$, $N = 0~\text{and}~2 $. Here the
singlet-triplet mixing is close to 40\% which makes the singlet
lines easily detectable. Additionally, for $v=31$ the ``s1" and
''d1" components are shifted to lower values than expected from
the reference spectrum $v=28$ due to repulsion by the singlet
component on the high energy side. In the case of $v=33$ the
singlet component pushes the ``s1" and ''d1" in the  opposite
direction. Our calculations show that even the chosen energy
reference ``s2" starts to show mixing, indicated by its reduced
triplet character of 0.98. In parallel to the singlet-triplet
mixing, the quantum number $I$ loses its meaning as well, e.g. the
level ``s1" of $v=33$ has an expectation value for $I$ of 2.75
instead of 3, due to a significant contribution of $I=2$ from the
singlet state. Comparing the stick spectrum line positions with
observed lines in Fig.\,\ref{fig:v28to33_1g} we note the very nice
agreement between experiment and theory. We want to mention that
we measured the singlet levels at the predicted positions from
first calculations. This emphasizes the predictive power of the
presented model.

The fact that we observe singlet-triplet mixing for relatively
deeply bound levels with binding energies of a few hundred
GHz$\times h$ is already quite interesting. In addition, it
provides valuable information for fixing the energy position of
the triplet levels with respect to the singlets with high
precision. This is especially important for the large body of data
of the singlet system, see section \ref{ccmodel}, which was
obtained with no connection to the triplet state \cite{seto:2000}.

\subsection{Franck-Condon overlap}
\begin{figure}[htbp] \centering
\includegraphics{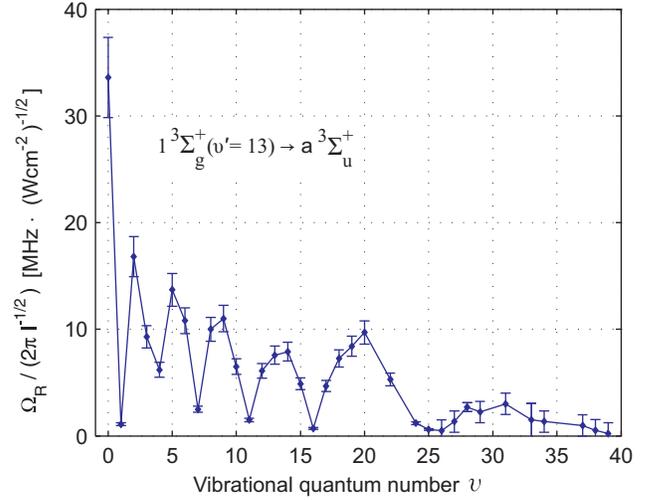}
\caption{Normalized transition matrix element $\Omega_2 /(2\pi
\sqrt{I_2})$ between the excited level $\qme$ ($\vib'=13, 1_g$
character) in the $\tripletex$ potential and the ``s1" level with
vibrational quantum number $v$ in the lowest triplet potential
$\triplet$. } \label{fig:FranckCondon}
\end{figure}
When scanning over all vibrational levels, the transition matrix
elements for the transition from $\qme$ to $\qmv$ are not constant
but oscillate as a function of $v$. This oscillation is mainly due
to variations in the Franck-Condon overlap between the $\qme, \
\vib'=13$ vibrational wave function and the vibrational wave
functions $\qmv$ of the $\triplet$ potential. Figure
\,\ref{fig:FranckCondon} shows the normalized transition matrix
element, $c_2=\Omega_2 (v) / (2\pi\sqrt{I_2})$, where $\Omega_{2}$
is the Rabi frequency and $I_{2}$ is the intensity of laser 2.
$\Omega_{2}$ is determined from the measured width of the dark
resonance \cite{Method,Win07}.  For these measurements we used the
excited level $1_g$ as $\qme$ and the ``s1" levels as $\qmv$ of
the $\triplet$ state. The transition matrix element $c_2$ varies
from about\, \mbox{($0.2$ to $33$)\,MHz/$\sqrt{\mbox{Wcm}^{-2}}$}.
In terms of a dipole moment, $\langle e r \rangle = \Omega_2 (v) /
\sqrt{I_2} \times \hbar \sqrt{\epsilon_0 c/2}$, this corresponds
to ($0.05$ to $8.0$)$\times10^{-30}$\,Cm.

In comparison, the amplitude for the transition between $\qmf$ and
$\qme$ is $\Omega_{1}/\sqrt{I_{1}}= $0.4\,MHz/
$\sqrt{\mbox{Wcm}^{-2}}$. We determine $\Omega_1$ from resonant
excitation, measuring how quickly $\qmf$ molecules are lost for a
given laser intensity $I_1$. The data in Fig.
\ref{fig:FranckCondon} can be used to fix the position of the
$\tripletex$ potential relative to the $\triplet$ potential
applying the Franck-Condon approximation.

\section{Conclusion and outlook}
\label{conclusion}
In this paper we demonstrate high resolution
dark state spectroscopy of ultracold $^{87}$Rb$_2$ molecules. We
are able to resolve the vibrational, rotational, hyperfine and
Zeeman structure of the lowest $\triplet$ potential with an
absolute accuracy as high as 30\,MHz. We find that the hyperfine
structure is weakly dependent on the vibrational level, which also
influences predictions of Feshbach resonances if one would like to
reach accuracies on the order of 0.1 G.

By optimizing mainly the triplet Born-Oppenheimer potential we
obtain a model that can quite accurately predict all vibrational
levels in the $\triplet$ and $\singlet$ potentials for different
ranges of rotational quantum numbers. After reassigning recent
data from Fourier transform spectroscopy \cite{Be09} we extend the
range of applicability to rotational states as high as $N$ = 70
for the $\triplet$ potential. Because of this reassignment also
the molecular parameters like the equilibrium internuclear
separation $R_e$ or the dissociation energy $D_e$ have changed
significantly compared to values reported in \cite{Be09}. The new
values are given in Table \ref{pota}. For rotational levels with
$N\leq4$ the model calculations for triplet levels of any
vibrational level should have a precision similar to that of our
measurements, \emph{i.e.} about 30 MHz. For higher $N$ this
precision will increasingly degrade resulting from the reduced
accuracy of the data from \cite{Be09} of about 300 MHz for $N=70$.
Compared to \cite{seto:2000} the potential of the ground state
$\singlet$ is significantly improved close to the atomic asymptote
by including data on the mixed singlet-triplet levels of this
study and data on Feshbach resonances from various other sources.
We recommend the derived potential function for further use. It
can predict the deeply bound levels with an accuracy (about 50
MHz), comparable to that of the Fourier transform spectroscopy in
\cite{seto:2000}. The asymptotic levels are accurate on the order
of a few MHz or better as their position is determined by the
precisely measured location of Feshbach resonances.
\begin{table}
\caption{Scattering lengths (in Bohr radius, $a_0=0.5292 \times
10^{-10}$m). $a_{lowest}$ gives the scattering length of the
energetically lowest hyperfine state at $B_z$ = 0.}
\label{tab:scat}
\begin{tabular}{|r|r|r|r r|} \hline
isotope & {$a_{singlet}$} & {$a_{triplet}$} & { $a_{lowest}$}& ${\rm Rb}+{\rm Rb}$\\
    &  &  &  & $(f,m_f) + (f,m_f)$\\
\hline
    $87/87$ &     $90.35$ &     $ 99.04 $ &  $ 100.36 $&$ (1,1)+(1,1)$   \\
    $85/85$ &     $2720$ &   $ -386.9 $ &       $-460.1  $&$  (2,2)+(2,2) $   \\
    $87/85$ &   $ 11.37 $ &    $ 201.0 $ &      $ 229.4 $&$  (1,1)+(2,2) $   \\
\hline
\end{tabular}
\end{table}
We improved the description of the large set of $^{87}$Rb Feshbach
resonances as compared to ref. \cite{marte:2002} where the
original calculation was based only on a few selected resonances
and a derived asymptotic form of the two potentials. In
\cite{marte:2002} deviations as high as 2\,G appeared and for
example the resonances of asymptote $f_A=1,m_A=0 + f_B=1,m_B=1$
were all calculated systematically too high. In the present model
these discrepancies disappear. The average deviation over all 46
resonances, given in \cite{marte:2002, roberts:2001, erhard:2003,
papp:2006}, is about 0.15\,G \cite{reass}. If one removes the
R-dependence of the hyperfine coupling as given in
Eq.\,\ref{eq:hfs}, deviations in the order of 0.5\,G appear
varying according to the different $f$ levels which correlate to
the Feshbach resonances. To improve the overall fit, one probably
needs to include many-body effects of the collision process for
detecting the resonances. The overall improvement of the long
range behavior also allows the whole set of scattering lengths for
the three isotopologues to be calculated. Table \ref{tab:scat}
gives a selection of these calculations. The error of these data
will be in the last digit shown. These values are consistent with
earlier publications but more precise and in general internally
consistent between the isotopologues.

Our model will be valuable for our planned collision experiments
of ultracold molecules which have to be prepared in well defined
quantum states. The singlet-triplet mixing that we observe is of
interest for a proposed precision experiment to measure the time
dependence of the electron to proton mass ratio \cite{DeM08} (see
also \cite{Ch09,Ze08}). The precision of future measurements would
be mainly limited by the accuracy of the wavemeter and could be
improved by one or two orders of magnitude. Such a high precision
could be used, e.g. to test the fundamental limits of the
coupled-channel model, where we assume the validity of (1) the
Born-Oppenheimer approximation, (2) using Zeeman terms with atomic
parameters only, (3) a limited functional dependence in $R$ of the
hyperfine and spin-spin interaction and (4) neglecting quadrupole
hyperfine coupling.

\begin{acknowledgments}
We thank Marius Lysebo and Leif Veseth for model calculations to
characterize the excited states $\qme$. We thank Matthias Gerster
for studying the simple Hamiltonian in Eq. (1). We acknowledge
helpful predictions from Christiane Koch on the Franck-Condon
overlap for the $\qme - \qmv$ transition. This work was supported
by the Austrian Science Fund (FWF) within SFB 15 (project part
17). E.T. thanks the cluster of excellence ``QUEST" for support
and the Minister of Science and Culture for providing a Niedersachsenprofessur.
\end{acknowledgments}

\smallskip
$^*$ These authors both contributed equally to this work.


\begin{thebibliography}{99}
\bibitem{width} The width of the dark resonance scales as
$\Omega_2$ if laser 2 is fixed on resonance and laser 1 is
scanned.
\bibitem{expectval} In general, we define expectation values for the quantum
numbers of an operator as $\sum_i \lambda_i |\langle  e_i | \Psi
\rangle |^2$, where $\lambda_i$ are the eigenvalues of the
operator, $e_i$ are corresponding normalized basis vectors, and $|
\Psi \rangle$ is the state.
\bibitem{Method}
In brief, using a three level model, where the Rabi frequency
$\Omega_2$ is the only free parameter, we fit a line shape to the
dark resonance \cite{Win07}.
\end{thebibliography}
\end{document}